\documentclass[aps,prb,twocolumn,floatfix,footinbib,showpacs,superscriptaddress]{revtex4-1}
\usepackage{graphicx}
\usepackage{amsfonts,amsmath,amssymb}
\usepackage{amsthm}
\usepackage{dsfont,bm}
\usepackage{color}
\usepackage{soul} 
\usepackage{amsbsy}
\usepackage[colorlinks=true,linkcolor=blue,pagecolor=blue,filecolor=blue,menucolor=blue,urlcolor=blue,citecolor=blue,anchorcolor=blue]{hyperref}%
\usepackage{sidecap}
\usepackage{txfonts}
\usepackage{amsbsy}
\usepackage{times} 
\usepackage{dcolumn}

\newcommand{\ffsf}{F$_1$F$_2$SF$_3\;$}
\newcommand{\ffsff}{F$_1$F$_2$SF$_3$F$_4\;$}

\newcommand{\ffsffz}{F$_1$F$_2$SF$_3$F$_4$}

\newcommand{\fsf}{F$_1$SF$_2\;$}
\newcommand{\sff}{SF$_1$F$_2\;$}

\newcommand{\fsfz}{F$_1$SF$_2$}

\begin{document}

\title{Half-Metallic Superconducting Triplet Spin MultiValves} 

\author{Mohammad Alidoust}
\affiliation{Department of Physics, K.N. Toosi University of Technology, Tehran 15875-4416, Iran}
\author{Klaus Halterman }
\affiliation{Michelson Lab, Physics Division, Naval Air Warfare Center, China Lake, California 93555, USA}

\date{\today} 
\begin{abstract}
We study  spin switching effects in finite-size superconducting
multivalve structures. We examine \ffsf and \ffsff hybrids where a
singlet superconductor (S) layer is sandwiched among ferromagnet (F)
layers with differing thicknesses and magnetization orientations. Our
results reveal a considerable number of experimentally viable spin valve
configurations that lead to on-off switching of the superconducting state.
For S widths on the order of the superconducting coherence length $\xi_0$,
non-collinear magnetization orientations in
adjacent F layers
with multiple spin-axes leads to a rich variety of triplet spin-valve effects. Motivated by recent experiments, 
we focus on
samples where magnetizations in   
the F$_1$ and F$_4$ layers exist in a 
fully spin-polarized half metallic phase,
and calculate
the superconducting transition temperature, spatially and energy resolved density of states, and the spin-singlet and spin-triplet superconducting correlations.
Our
findings demonstrate that superconductivity in these devices can be
completely switched on or off over a wide range of magnetization misalignment
angles due to the generation of equal-spin and opposite-spin triplet pairings.  
\end{abstract}
\maketitle

\section{introduction}
Over the last two decades,
the interplay of superconductivity and ferromagnetism has fueled  interest in exploring 
ferromagnet (F) and superconductor (S)
hybrid structures for low temperature spintronic 
applications\cite{fsf2,Eschrig2015RPP,Buzdin2005,first,blamire,Beenakker2013ARCM,Nayak2008RMP}. One intriguing consequence of this interplay is the creation of spin-triplet Cooper pairs that 
was predicted theoretically\cite{fsf2,Buzdin2005,first}. To confirm the generation of these unconventional pairs, 
much progress
 has been made so far, both theoretically and 
experimentally \cite{sfs1,sfs2,sfs3,sfs4, halt2,Moor,Khaydukov,Anwar,Zdravkov,Antropov,Halterman2007,silaev,tanaka,Pugach,Enderlein,Iovan,Cayao,Tkachov,valve1,bergeret1,alidoust1,alidoust2,Pugach}. One of the
  first signatures of the
  existence of spin-polarized superconducting correlations was observed in a planar half-metallic Josephson junction \cite{Keizer2006}. Since a half-metal supports only one spin direction, it was concluded that the supercurrent should be carried by an equal-spin triplet channel. 

The two kinds of basic spin valves that have been mainly studied both
experimentally and theoretically are \fsf and \sff based structures   
\cite{fsf1,fsf2,fsf3,fsf41,fsf5,fsf6,fsf7,sff1,sff2,prsh1,prsh2,prsh3,sfsf,jap_al,andeev,sff3}. These
systems offer simple and controllable platforms that can reveal 
signatures of spin triplet superconducting correlations. If differing 
ferromagnetic materials, constituting the left and right F layers are
chosen properly, they respond to an external
magnetic field in different ways,  providing  active control
of the magnetization misalignment angles through variations in the intensity and
direction of an external magnetic field. It was shown that the
superconducting transition temperature \cite{singh,half,bernard1,gol1,Mironov}
and density of states \cite{bernard2,zep,zep2,zep3,Anwar} reveal
prominent signatures of the long-ranged spin-triplet  superconductivity\cite{fsf2} as a function of magnetization misalignment. 
Nevertheless, a direct and clear observation of the equal-spin triplet pairings in superconducting hybrids is still elusive.

In a recent experiment\cite{singh} involving a
SF$_1$F$_2$ spin
valve, it was observed that the
superconducting critical temperature $T_c$ in 
{\rm MoGe}-{\rm Ni}-{\rm Cu}-{\rm CrO}$_2$, where F$_2$ is the half-metallic compound, CrO$_2$,    can 
have a variation as high as $\Delta T_c\sim 800$mK when varying  the magnetization
misalignment angle. This order of $T_c$ variation is much
larger compared to when standard ferromagnets are used 
[i.e., {\rm CuNi}-{\rm Nb}-{\rm CuNi} \cite{fsf5}, {\rm
  CoO}$_x$-{\rm Fe}1-{\rm Cu}-{\rm Fe}1-{\rm Pb}\cite{sff2}, {\rm
  Co}-{\rm Cu}-{\rm Py}-{\rm Cu}-{\rm Nb} \cite{sff3}, and {\rm Co}-{\rm Nb}-{\rm Co}-{\rm Nb}\cite{sfsf}],
 albeit using a relatively  strong {\it out-of-plane} magnetic
field of $H\sim 2$T. This was consistent  with theory that demonstrated 
the largest variations in 
$T_c$ of a  \sff spin valve \cite{half} occurred when  F$_2$ is a half-metal, rather
than a  ferromagnet with a smaller exchange energy\cite{bernard1}. 
A  very recent experiment\cite{bernard1} involving the half metal  
La$_{0.6}$Ca$_{0.4}$MnO$_3$, consisted of a {\rm
  LCMO}-{\rm Au}-{\rm Py}-{\rm Cu}-{\rm Nb} stack under the influence of a much weaker 
{\it in-plane} magnetic field of $H\sim 3.3$mT to rotate
 the magnetization. This configuration  
demonstrated a slight improvement with $\Delta T_c\sim 150$mK compared 
 to experiments involving  standard ferromagnets that yielded $\Delta T_c \sim 50$mK\cite{sff2}, and
$\Delta T_c \sim 120$mK\cite{sff3}. 
Since strong variations in $T_c$ can be representative of singlet superconductivity weakening
and its conversion to the triplet channel, 
it is of fundamental
interest to create spin valve structures with the largest variations in $T_c$ by a \textit{weak} external magnetic field. 
On the one hand, it can provide unambiguous signatures of the presence of equal-spin triplet
correlations under magnetization rotation.
On the other hand, by restricting  the magnetization variations to reside in-plane, the overall magnetization
state in the ferromagnet can be manipulated with much weaker fields. This low-field magnetization control considerably  
 increases device reliability and provides an effective spin switch for technological purposes. 

\begin{figure}
\includegraphics[width=8.5cm, height=5.60 cm]{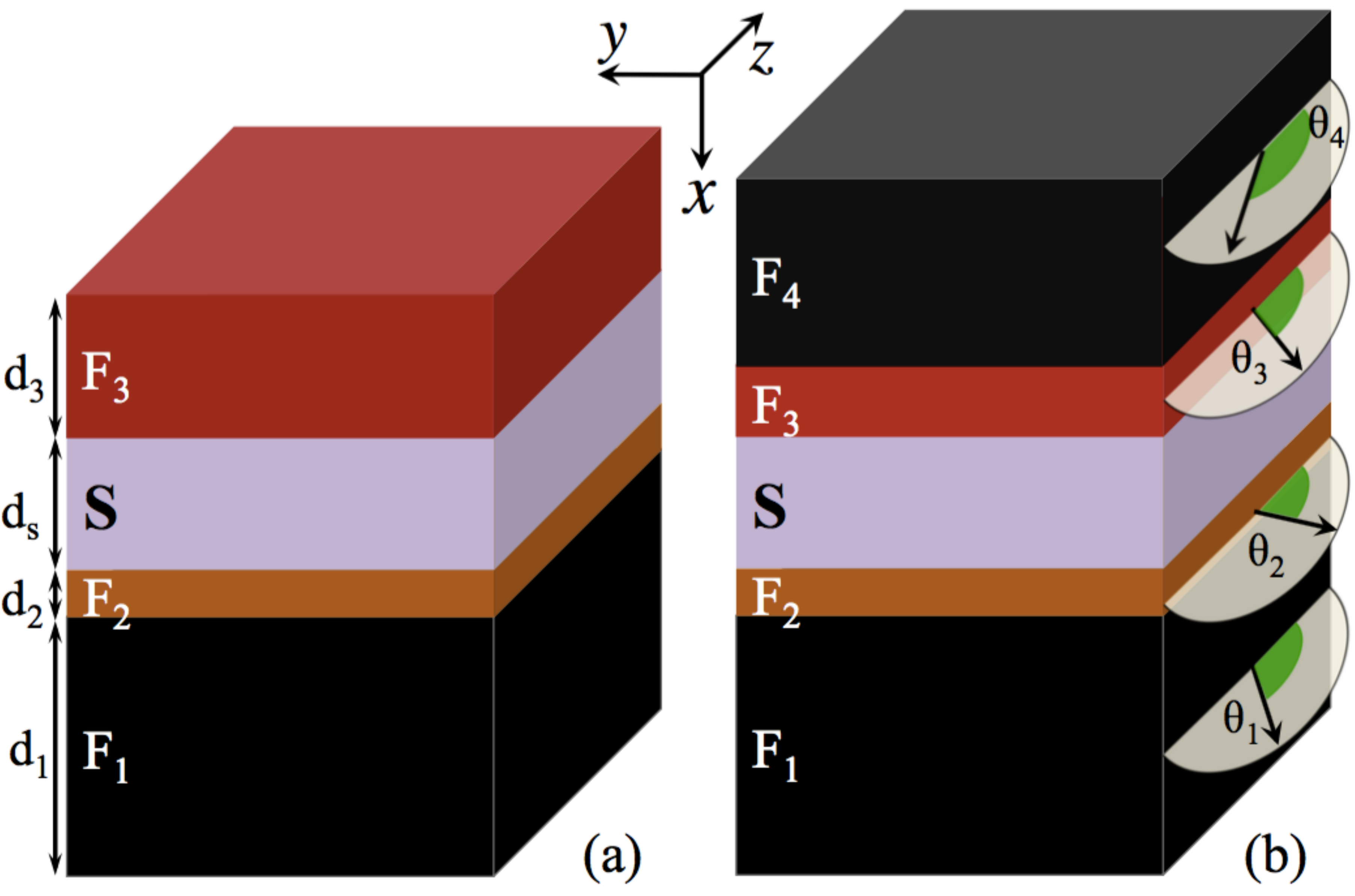}
\caption{Schematic of the superconducting triplet spin multivalves: (a) \ffsf and (b) \ffsff structures. 
The interfaces reside in the $yz$ plane and the thickness of a layer $i$=1-4 is marked by d$_i$. 
The magnetization of the F layers is shown by an arrow that can make arbitrary angles with respect to the $z$ axis marked by $\theta_i$.   }
\label{fig1}
\end{figure}
In this paper, we introduce the \ffsf and \ffsff multivalves
(depicted in Fig.~\ref{fig1}), 
where the two outer F layers are half-metallic. The presence of the
additional ferromagnetic layers
amplifies the
singlet-triplet conversion process, leading to a larger spin valve effect
compared to
standard spin valves described above. 
We show this by considering a wide range of
layer thicknesses and magnetizations to
achieve a broad critical temperature variation
as the magnetization in one of the F layers rotates.
As mentioned earlier, the transition temperature of 
a spin valve system\cite{singh,half,bernard1,gol1} and the density of
states\cite{bernard2,zep,zep2,zep3} are two 
experimental 
quantities that can determine the degree in which 
triplet
pair conversion takes place for a given
device. Our investigation is within the ballistic regime using 
a microscopic self-consistent formalism
that can accommodate the large variation in energy scales present in 
the problem. In particular, the multivalve structures considered here contain conventional
ferromagnets with relatively weak exchange fields ($h/E_F\ll1$) in addition
to the surrounding half metals ($h/E_F=1$).
Our results
demonstrate that, stemming from 
nontrivial interlayer spin-valve effects,
superconductivity can be
effectively switched on or off over a wide range of relative 
magnetization misalignment
angles.
We note that this feature
is absent or occurs 
to a 
limited extent in the previously studied\cite{switch, half} individual \fsf and \sff spin-valves.  
We complement  the $T_c$ studies  by 
investigating various pair correlations and the corresponding local density of states,
where the emergence of a zero-energy peak is associated with the 
presence of equal-spin triplet  correlations.  

The rest of the paper is organized as follows. We first present an overview of the
theoretical framework used. In Sec.~\ref{tc} we study
the superconducting 
transition temperature for two multivalve configurations. 
In Sec.~\ref{dos} we present
the corresponding local density of states, paying particular attention to the 
DOS at zero energy. 
Lastly, we present  the singlet and triplet superconducting correlations generated and discuss 
their behavior in Sec.~\ref{corr}. Finally, we give concluding remarks in Sec.~\ref{con}.

\begin{figure*}
\centering
\includegraphics[width=11.3cm]{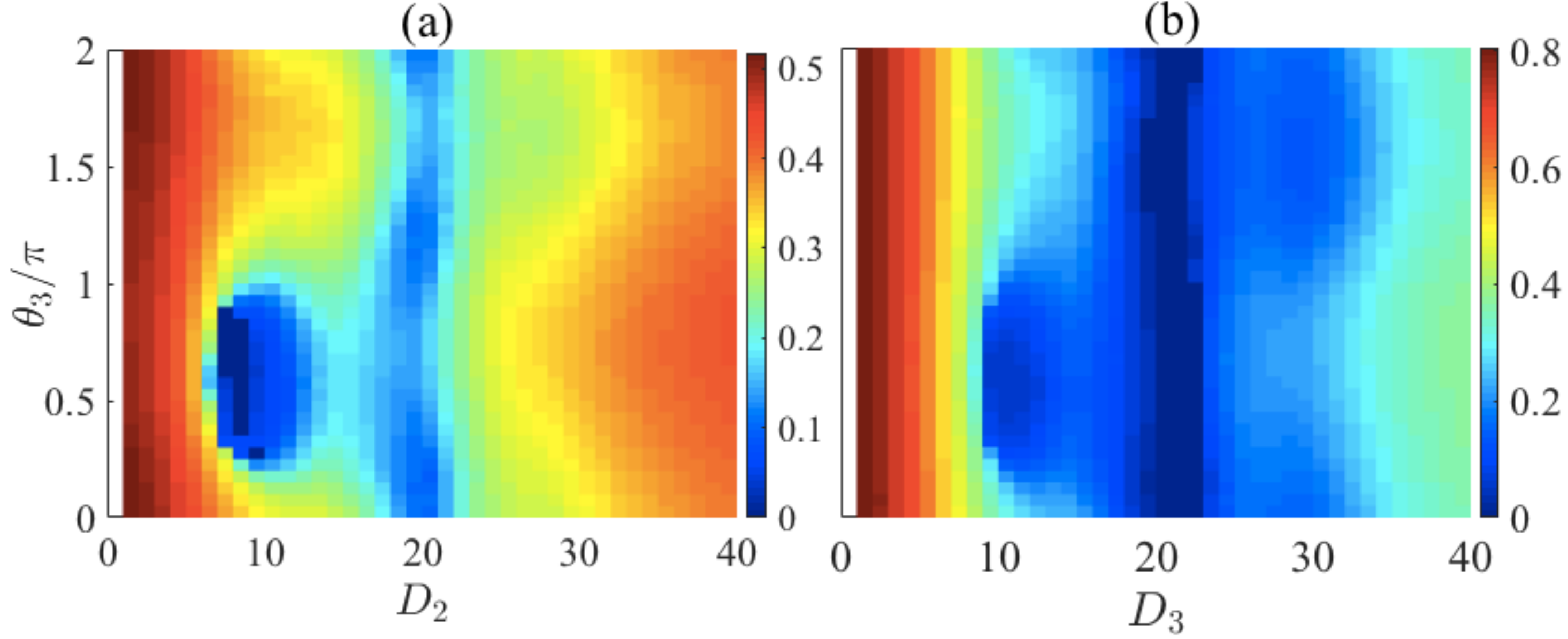}
\centering
\caption{The normalized transition temperature as a
  function of magnetization rotation in F$_3$ layer of a \ffsf spin
  multivalve. In panel (a) the thickness of F$_2$ layer changes while
  in (b) the thickness of F$_3$ varies.}
\label{ffsf_b_d}
\end{figure*}

\section{theory and results}\label{t&r}

To accurately describe the ballistic multivalve configurations displayed in Fig.~\ref{fig1},
we solve the spin-dependent Bogoliubov-de Gennes (BdG) equations in a fully self-consistent\cite{halter1} manner. In this formalism we denote the quasiparticles' energy and their associated probability amplitudes by 
$\epsilon_n$, $u_{n\sigma}$, and $v_{n\sigma}$ ($\sigma=\uparrow,\downarrow$), respectively:
\begin{align}
&\begin{pmatrix}
{\cal H}_0 -h_z&-h_x+ih_y&0&\Delta({x}) \\
-h_x-ih_y&{\cal H}_0 +h_z&\Delta({x})&0 \\
0&\Delta^*({x})&-({\cal H}_0 -h_z)&-h_x-ih_y \\
\Delta^*({x})&0&-h_x+ih_y&-({\cal H}_0+h_z) \\
\end{pmatrix}\times\nonumber \\
&
\begin{pmatrix}
u_{n\uparrow}\\u_{n\downarrow}\\v_{n\uparrow}\\v_{n\downarrow}
\end{pmatrix}
=\epsilon_n
\begin{pmatrix}
u_{n\uparrow}\\u_{n\downarrow}\\v_{n\uparrow}\\v_{n\downarrow}
\end{pmatrix},
\label{bogo}
\end{align}
in which $\Delta({x})$ represents
the pair potential,  calculated 
self-consistently as shown below [Eq.~(\ref{sc})].
For the in-plane magnetization rotations considered here,
the components of the exchange field $\bm h$ in each of the F layers take the form:
${\bm h} = (h_{x}(x),h_{y}(x),h_{z}(x))$,
so that the exchange field which vanishes in the S layer, can in general vary between the ferromagnet layers.
The free-particle Hamiltonian ${\cal H}_0(x)$ is defined as,
\begin{equation}
{\cal H}_0(x)\equiv\frac{1}{2m}\left(-\frac{d^2}{dx^2}+k_y^2+k_z^2
\right)-E_F + U(x),
\end{equation}
where $E_F$ denotes the Fermi energy, and $U(x)$ is a spin-independent
scattering potential.
The multivalve layers  
in Fig.~\ref{fig1}, 
are translationally invariant in the $yz$ plane and hence quasiparticle motion in this plane
is appropriately  described by the good quantum numbers $k_y$ and $k_z$.
For this reason, all spatial variations take place in the $x$ direction, and the system is considered quasi-one dimensional.

Ferromagnetism and superconductivity are two 
competing types of ordering and their simultaneous existence in space results in nontrivial
spatial profiles for the pair potential $\Delta(x)$. Therefore, to account properly for proximity effects, 
it is necessary to obtain the pair potential in a self-consistent manner via:
\begin{align} \label{sc}
\Delta(x) = \frac{g(x)}{2}{\sum_{n}^{\omega_D}}
[u_{n\uparrow}(x)v^*_{n\downarrow} (x)+
u_{n\downarrow}(x)v^*_{n\uparrow} (x)]\tanh\left(\frac{\epsilon_n}{2T}\right),
\end{align}
where $g(x)$ is the pair coupling constant that is nonzero solely
inside the superconductor layer and the sum above is restricted to
the quantum states with positive energies below
the 
Debye energy cutoff $\omega_D$.

To compute the transition temperature, we use
the fact that $\Delta(x)/\Delta_0\ll1$ close to the critical temperature,
where $\Delta_0$ is the bulk pair potential at zero temperature, $T=0$.  
In this regime,  
the
self-consistency equation [Eq.~(\ref{sc})] can be linearized near the
transition through a perturbative expansion of the quasiparticle amplitudes and energies
and retaining terms to first ordfixer. 
As part of the linearization process, the pair potential $\Delta(x)$ and 
quasiparticle amplitudes 
are  Fourier expanded in a sine-wave basis. 
For example,  the zeroth-order
wavefunctions  are written,  
$u^0_{n \sigma}=\sqrt{2/d}\sum_p u_{np}^\sigma\sin(k_p x)$
and $v^0_{n \sigma}=\sqrt{2/d}\sum_p v_{np}^\sigma\sin(k_p x)$,
where $k_p=p\pi/d$.
Using standard perturbation theory techniques described elsewhere,\cite{tcpap,fsf5}
we arrive at the following matrix eigenvalue problem:
\begin{align} \label{eig}
\Delta_i=\sum_q J_{i q}\Delta_q,
\end{align}
where the $\Delta_q$
are the first-order expansion coefficients for $\Delta(x)$, and $J_{iq}$ are the
matrix elements 
written entirely in terms of zeroth-order quantities:
\begin{align}
J_{iq} = \frac{g N_0}{8\pi k_F d}\int d\epsilon_\perp & \sum_n \sum_m
\Biggl\{ \frac { {\cal F}_{qnm} {\cal F}_{inm}}{\epsilon_{n}^u-\epsilon_m^v}\tanh\left(\frac{\epsilon_n^u}{2T}\right)\nonumber \\
&+ \frac { {\cal F}_{qmn} {\cal F}_{imn}}{\epsilon_{n}^v-\epsilon_m^u}\tanh\left(\frac{\epsilon_n^v}{2T}\right)\Biggr\},
\end{align}
where $\varepsilon_{\perp}=1/(2m)(k_y^2+k_z^2)$ is 
the quasiparticle  kinetic energy 
in the transverse direction, $N_0$ is the density of states at the Fermi energy,
and $\varepsilon_n^{u,v}$ are the unperturbed zeroth-order energies.
The zeroth-order quasiparticle amplitudes ($u^0_{n \sigma}$, $v^0_{n \sigma}$) 
and corresponding energies ($\varepsilon_n^{u}$,$\varepsilon_n^{v}$) are found by solving 
Eq.~(\ref{bogo}) with  $\Delta(x)=0$.
We also define,
${\cal F}_{qnm}=\pi \sqrt{2d} \sum_{p,r}  { K}_{qpr} (u_{mp}^\uparrow  v_{nr}^\downarrow+u_{mp}^\downarrow  v_{nr}^\uparrow)$,
where 
\begin{equation}
gK_{q p r}=({2}/{d})^{3/2}\int_0^d dx g(x) \sin(k_q x) \sin(k_p x) \sin(k_r x).
\end{equation}

Experimentally accessible information regarding the quasiparticle spectra is 
contained in the local density of single-particle excitations in the system. 
This includes  zero-energy signatures in the form of peaks\cite{bernard2,zep,zep2,zep3} in the density of states (DOS), 
which can reveal the emergence
 of equal-spin triplet pairings  in either the ferromagnet\cite{zep} or superconductor\cite{half} regions. 
The total local density of states, $N(x,\epsilon)$, includes contributions from 
both the spin-up and spin-down quasiparticle states:
$N(x,\epsilon)={N}_{\uparrow}(x,\epsilon)+{N}_{\downarrow}(x,\epsilon)$, where,
\begin{equation}\label{dos}
{N}_\sigma(x,\epsilon)
=-\sum_{n}
\Bigl\lbrace|u_{n\sigma}(x)|^2
 f'(\epsilon-\epsilon_n)
+|v_{n\sigma}(x)|^2
 f'(\epsilon+\epsilon_n)\Bigr\rbrace,
\end{equation}
in which $f'(\epsilon) = \partial f/\partial \epsilon$ is the derivative
of the Fermi function.

In order to study the
various superconducting correlations that can arise, 
we define\cite{Halterman2007,halter_trip2} the triplet pair amplitudes in terms of the field operators:
\begin{subequations}
\label{pa}
\begin{align}
&&{f_0}({x},t) = \frac{1}{2}\left[\left\langle
\psi_{\uparrow}({x},t) \psi_{\downarrow} ({x},0)\right\rangle+
\left\langle \psi_{\downarrow}({x},t) \psi_{\uparrow} ({x},0)\right\rangle\right],\\
&&{f_1}({x},t) = \frac{1}{2}\left[\left\langle
\psi_{\uparrow}({x},t) \psi_{\uparrow} ({x},0)\right\rangle
-\left\langle \psi_{\downarrow}({x},t) \psi_{\downarrow} ({x},0)\right\rangle\right], \\
&&{f_2}({x},t) = \frac{1}{2}\left[\left\langle
\psi_{\uparrow}({x},t) \psi_{\uparrow} ({x},0)\right\rangle
+\left\langle \psi_{\downarrow}({x},t) \psi_{\downarrow} ({x},0)\right\rangle\right].
\end{align}
\end{subequations}
where $t$ is the relative time in the Heisenberg picture.
If we consider the quantization axis fixed along the $z$ axis, the triplet amplitudes ($f_{0}(x,t)$,$f_{1}(x,t)$,$f_{2}(x,t)$)
 can be 
written in terms of the quasiparticle amplitudes:~\cite{Halterman2007,halter_trip2}
\begin{subequations}
\begin{align}
f_{0}(x,t) &=  \frac{1}{2}\sum_{n}
\left[u_{n \uparrow}(x) v^{\ast}_{n\downarrow}(x)
-u_{n \downarrow}(x) v^{\ast}_{n\uparrow}(x)
\right] \zeta_n(t), \label{f0} \\
f_{1}(x,t) & =-\frac{1}{2} \sum_{n}
\left[
u_{n \uparrow}(x) v^{\ast}_{n\uparrow}(x)
+u_{n \downarrow}(x) v^{\ast}_{n\downarrow}(x)
\right]\zeta_n(t), \label{f1} \\
f_{2}(x,t) & =-\frac{1}{2} \sum_{n}
\left[
u_{n \uparrow}(x) v^{\ast}_{n\uparrow}(x)
-u_{n \downarrow}(x) v^{\ast}_{n\downarrow}(x)
\right]\zeta_n(t), \label{f2}
\end{align}
\end{subequations}
where
$\zeta_n(t)$ is given by,
\begin{align}
 \zeta_n(t) = \cos(\epsilon_n t)-i\sin(\epsilon_n t) \tanh\left(\frac{\epsilon_n}{2 T}\right). 
\end{align}

\begin{figure*}
\includegraphics[width=15.cm, height=6.0 cm]{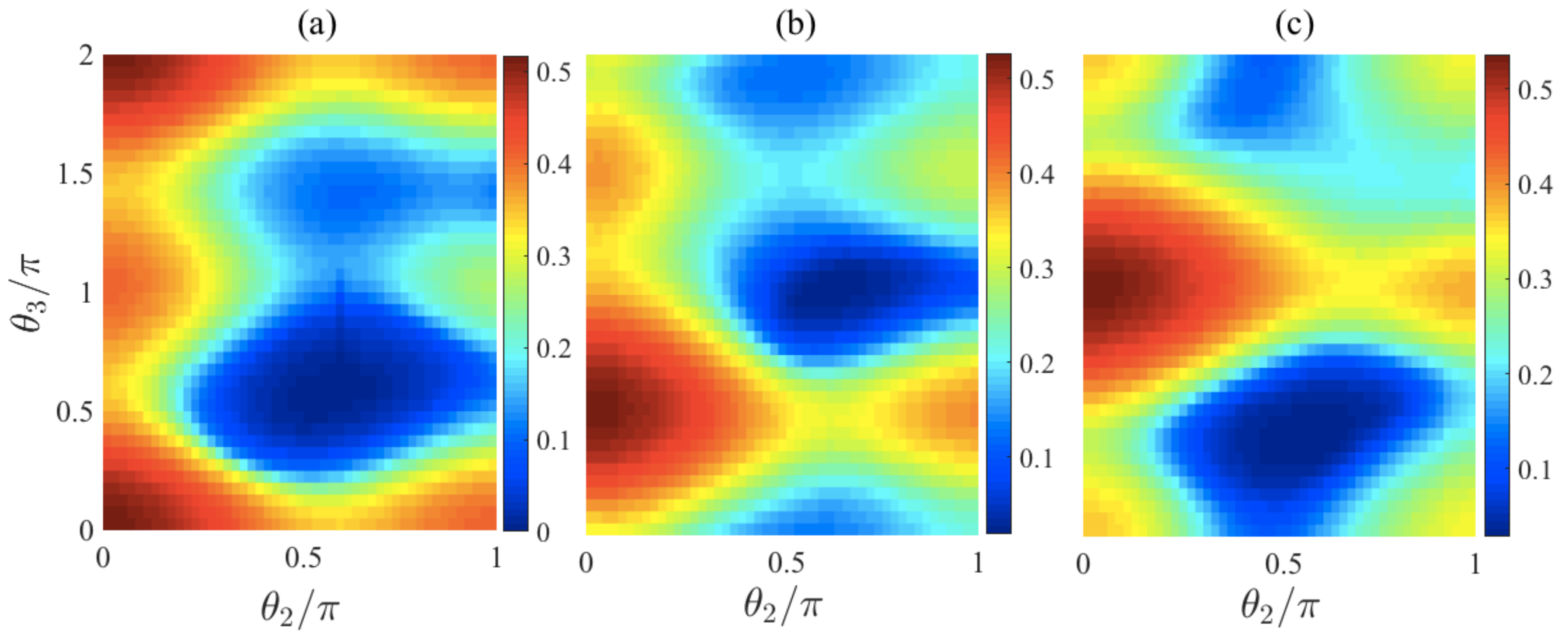}
\includegraphics[width=18.0cm, height=5.20 cm]{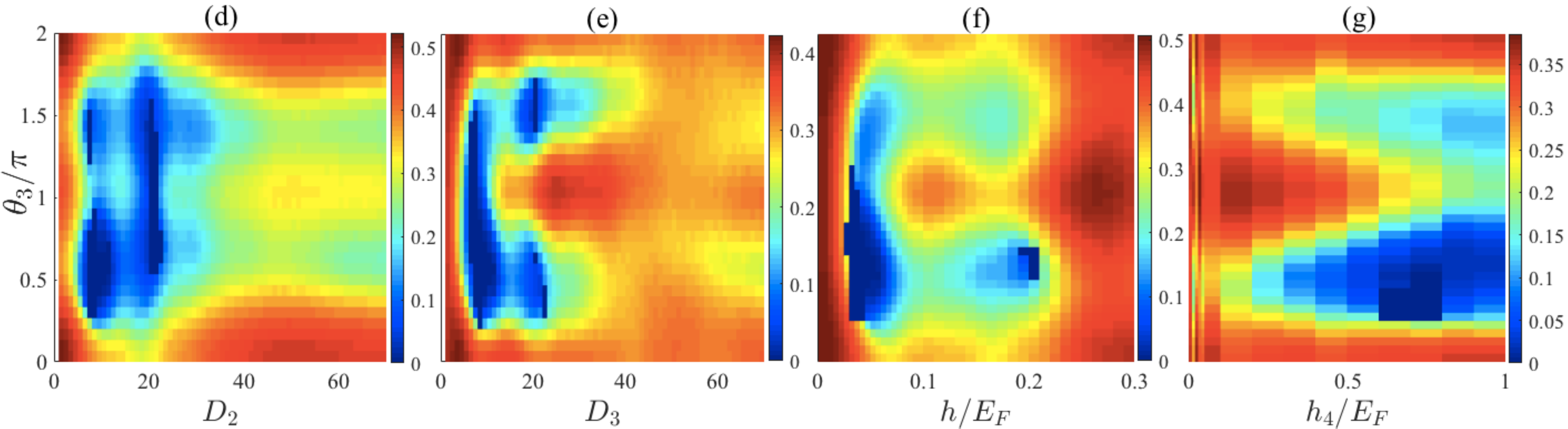}
\caption{The normalized transition temperature
  $T_c/T_0$ of a \ffsff structure where the outer magnetic
layers are in a half-metallic phase in panels (a)-(e). In panels (a)-(c) the critical
temperature is plotted as a function of $\theta_2$ and $\theta_3$, the
magnetization orientation in F$_2$ and F$_3$ layers. Here in (a), (b),
and (c),
we rotate the magnetization of F$_4$ so that
$\theta_4=0, 0.5\pi,$ and $\pi$, respectively. In panels (d) and (e),
the critical temperature is plotted as a function of magnetization
rotation in the F$_3$ layer and thicknesses of F$_2$ and F$_3$ layers:
$D_{2,3}$. Panel (f) shows $T_c/T_0$ vs $\theta_3$ and magnetization
strength in both F$_2$ and F$_3$. Here, we assume identical magnitudes
to the exchange fields, i.e., $h_2=h_3=h$. In (g), we plot $T_c/T_0$
vs the magnetization strength of F$_4$ and $\theta_3$ while $h_1=E_F$, supporting only one spin direction.} 
\label{ffsff_tc}
\end{figure*}

We consider an in-plane\cite{bernard1} 
Stoner-type exchange field for each
magnetic layer,
\begin{align}
\label{fields}
{\bm h}_i=h(0,\sin\theta_i,\cos\theta_i),
\end{align}
 where $h_{i}$ are the magnitudes of the exchange field in
each layer denoted by $i$, and $\theta_i$ are the angles that ${\bm h}_i$ make with
the $z$-axis (see Fig.~\ref{fig1}).
In situations where it is more convenient to align the quantization axis with
the local exchange field direction, we perform a
spin rotation using the transformations 
found in 
the Appendix.
The result is 
\begin{align}
f_0'&=f_0 \cos\theta_i 
+i\sin\theta_i f_2,\\
f_1'&=f_1,\\
f_2'&=\cos\theta_i f_2
+i\sin\theta_i f_0.
\end{align}

When presenting results, we normalize all lengths by the Fermi wavevector
$k_F$, e.g., $D_i=k_F d_i$, $X=k_F x$. We set the superconducting coherence length to the
normalized value of $k_F \xi_0=100$.  The half-metallic layers have
fixed widths $D_{1,4}=100$.
The quasiparticle energies are scaled by the bulk superconducting  gap
$\Delta_0$, and the critical temperature of a sample by $T_0$, the transition temperature of its  bulk counterpart.
The local DOS is normalized by the DOS of a normal metal.

\subsection{Transition Temperature}\label{tc}

In this section, we present results for the transition temperature $T_c$  
by solving the matrix eigenvalue equation in Eq.~(\ref{eig}). 
The procedure for identifying $T_c$ involves
varying the temperature $T$ and calculating the eigenvalues\cite{tcpap,fsf5}.
At the
transition temperature, the largest eigenvalue is unity,
while if $T>T_c$ all eigenvalues are less than unity.
In Fig.~\ref{ffsf_b_d}, the normalized transition temperature $T_c/T_{0}$
of the
 \ffsf multivalve configuration  is shown (see Fig.~\ref{fig1}(a)), where F$_1$ is a
half-metal (with $h_1=E_F$). 
For both panels we have set $D_1=100$, $D_S=250$, $\theta_1=0,
\theta_2=\pi/2$, and $h_2=h_3=0.05E_F$. In Fig.~\ref{ffsf_b_d}(a) we set
$D_3=10$ and plot
$T_c$ vs $\theta_3$ and $D_2$. 
In Fig.~\ref{ffsf_b_d}(b) the
thickness of the F$_2$ is fixed at $D_2=10$ and $T_c$ is plotted vs
$\theta_3$ and $D_3$. As seen, the critical temperature approaches
zero in the window $0.35\pi\lesssim\theta_3\lesssim 0.8\pi$ and
$D_2\sim 10$. The critical temperature shows a nonmonotonic  behavior vs
both $\theta_3$ and $D_2$ with two minima at $D_2\sim 10$ and $D_2\sim
20$. Similar trends are obtained when $D_3$ varies. Nonetheless,
Fig. \ref{ffsf_b_d}(b) shows an effectively stronger spin valve effect
with $T_c$ spanning the range 
$0<T_c/T_0<0.8$, compared to Fig.~\ref{ffsf_b_d}(a) which has
$0<T_c/T_0<0.5$. 
In addition, at $D_3\sim 20$, the critical
temperature is vanishingly small for all values of $\theta_3$. 
These
results demonstrate an effective spin switch that can turn superconductivity on
or off using a
multivalve \ffsf configuration with experimentally accessible parameters. 
\begin{figure}
\includegraphics[width=8.cm, height=10.0 cm]{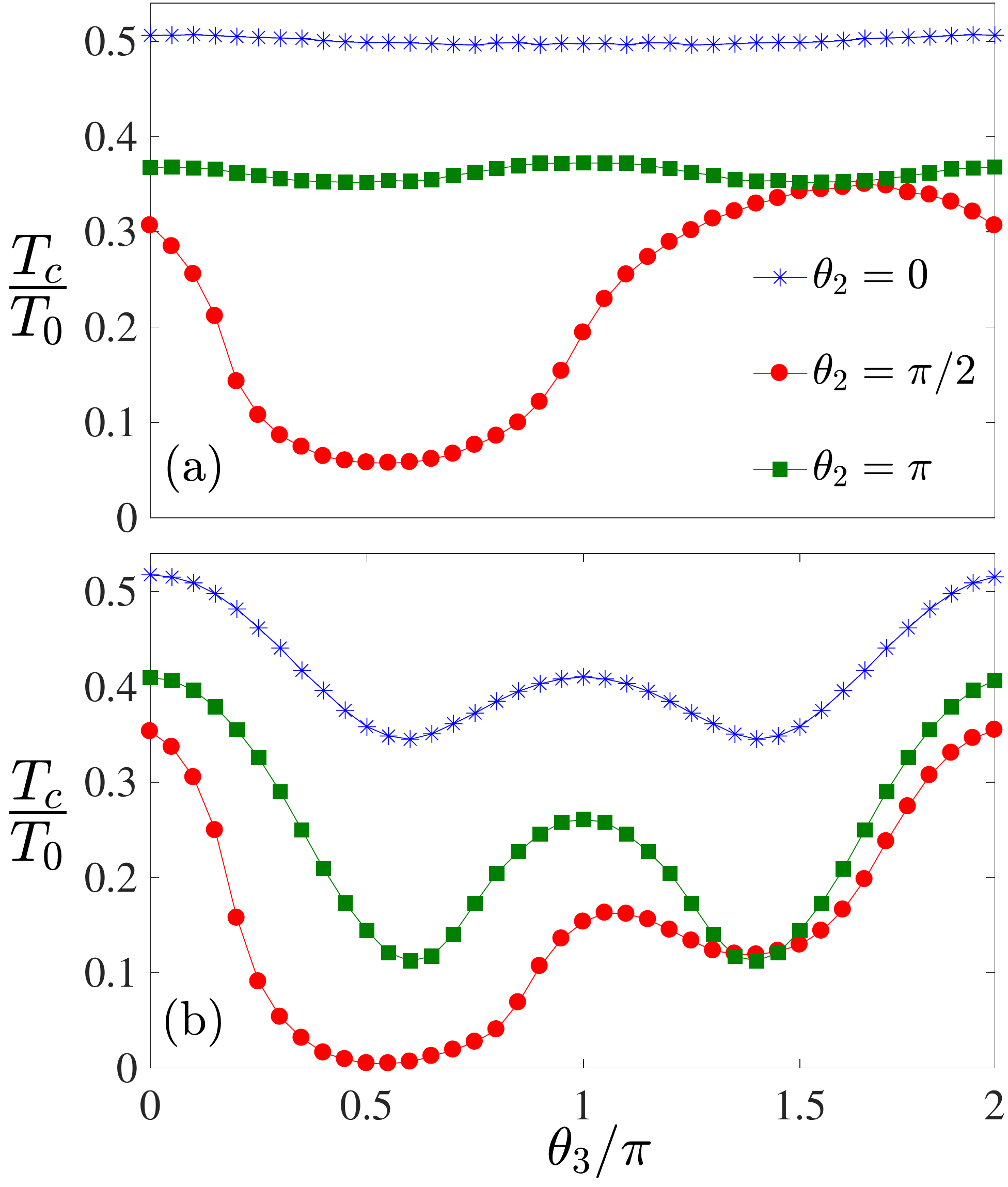}
\caption{The normalized transition temperature as a function of $\theta_3$, the magnetization rotation angle in $\rm F_3$.
Three different magnetization orientations in the $\rm F_2$ region are considered: 
$\theta_2=0, \pi/2,$ and $\pi$. 
Panel (a) illustrates the variation in the critical temperature of the $\rm F_1F_2SF_3$ structure, 
while (b) corresponds to a  $\rm F_1F_2SF_3F_4$ configuration.}
\label{fig:comp}
\end{figure}

\begin{figure*}
\includegraphics[width=18cm, height=9.0 cm]{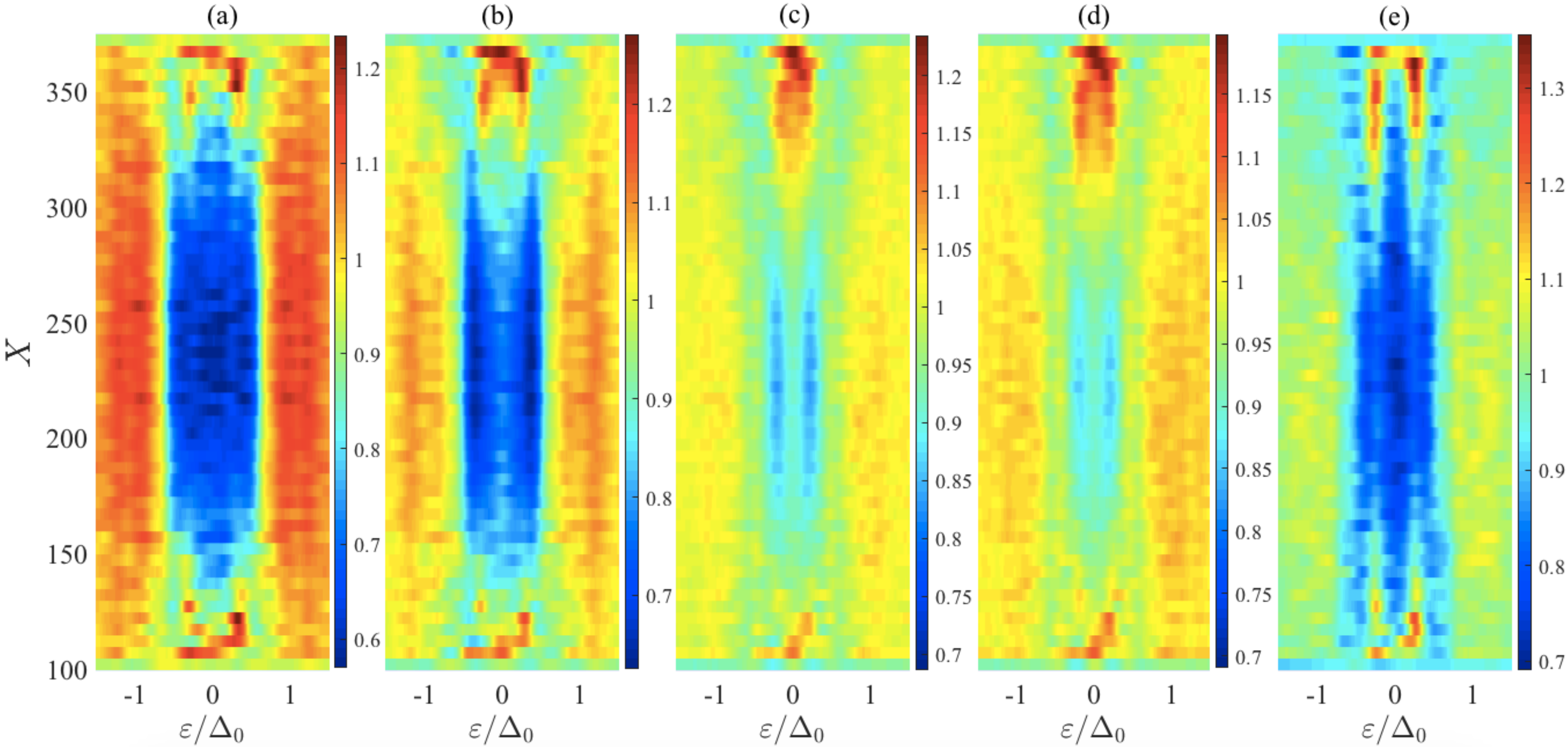}
\caption{The normalized spatial ($X\equiv k_F x$) and energy ($\varepsilon/\Delta_0$) resolved density
  of states of a ballistic \ffsff structure. The F$_{1,4}$ layers are in the
  half-metallic phase, and the magnetization in the F$_{1,2,4}$ layers is
  fixed along the $z$ axis. Each panel  (a)-(e) corresponds to  
  a different magnetization orientation, with $\theta_3=0,
  \pi/3.6, \pi/2, \pi/1.65$, and $\pi$, respectively. }
\label{dos_b}
\end{figure*}

Next, we incorporate an additional half-metallic layer, and consider the superconducting critical temperature of the
 \ffsff multivalve in Fig.~\ref{ffsff_tc}. 
 Having the
outer  $\rm F_1$ and $\rm F_4$ layers half-metallic with $h_{1,4}=E_F$ (and thus
only one spin band at the Fermi level)
 maximizes the generation of equal-spin
 triplet pairs.
 In Figs. \ref{ffsff_tc}(a), \ref{ffsff_tc}(b), and \ref{ffsff_tc}(c)
the normalized critical temperature $T_c/T_{0}$ is plotted as a function of
the magnetization orientation angles
in the
F$_{2,3}$ layers
$\theta_2$ and $\theta_3$. 
Here, the weaker inner ferromagnets have the exchange fields
$h_2=h_3=0.05 E_F$ and widths $D_2=D_3=10$.
The superconductor has $D_S=250$, corresponding to a relative thickness
$d_s/\xi_0=2.5$. 
In
Figs. \ref{ffsff_tc}(a), \ref{ffsff_tc}(b), and \ref{ffsff_tc}(c) the magnetization  in
the right
half-metal F$_4$ is set along (a) $z$, (b) $y$, and (c) $-z$, while
while the orientation of
the  left half-metal  F$_1$ is fixed along
the $z$ direction. As seen, in Fig.~\ref{ffsff_tc}(a), the critical
temperature is zero  at $\theta_{2,3}\sim \pi/2$. Thus,
at this magnetization configuration, the system transitions to
a normal resistive state  for $\it all$ temperatures.
By changing the magnetization alignment in 
 F$_4$, panels (b) and (c) demonstrate that regions of
 very low $T_c$ (blue regions)  shift
  to larger $\theta_2$ and
$\theta_3$. 
The critical temperature mappings in (a)-(c)
suggest that a large number of combinations of
magnetization alignments leads to effective 
spin switches with large critical temperature variations $\Delta T_c(\theta_i) = T_{c, {\rm max}}(\theta_i)-T_{c,{\rm min}}(\theta_i)$. 
In Fig.~\ref{ffsff_tc}(b), as seen, the minimum of $T_c$ exceeds zero and also
regions of very small $T_c$ occur over narrower angular ranges of
  $\theta_2$ and $\theta_3$. 

Next in Fig.~\ref{ffsff_tc} we investigate how the critical
temperature is modified when varying 
$\theta_3$ (we set $\theta_2=\pi/2$)  
and (d) the dimensionless
thickness $\rm D_2$ ($D_3=10$)
or (e) the thickness $\rm D_3$ ($D_2=10$).  
For certain ferromagnet widths
$D_{2,3} \sim 10$ and $20$,
the 
critical temperature 
is severely diminished for
a broad range of $\theta_3$. Indeed,
Fig. \ref{ffsff_tc}(d) shows that if the system is
in a superconducting state at $T=0$,
superconductivity is completely switched off
for orientations with 
$\theta_3\sim \pi/2$ and $\sim 1.5\pi$. 
This type of
switching effect occurs  over an extended angular range $0.5\pi \lesssim \theta_3 \lesssim
1.5\pi$ for $D_3\sim 20$, indicative of a strong spin valve effect in this regime. 
The same
feature appears in Fig. \ref{ffsff_tc}(e) except now the on-off superconductivity switching regime 
occurs at $D_3=10$ and broadens to $0.2\pi \lesssim \theta_3 \lesssim
1.5\pi$. 

The choice of relatively weak ferromagnets for $\rm F_2$ and $\rm F_3$
 generates  opposite-spin triplet correlations in those regions and a subsequent 
 conversion into spin-triplet pairs that can propagate within the half-metallic regions,
 modifying $T_c$.
 To identify the optimal ferromagnetic strengths, we show
 in Fig.~\ref{ffsff_tc}(f) 
the critical temperature 
vs the normalized magnetization strength in both
F$_2$ and F$_3$. For simplicity, we set $h_{2}=h_{3}=h$. 
It is evident that 
there is a broad region spanned by $h$  and $\theta_3$ 
in which $T_c\approx 0$, creating an effective superconductivity on-off switch.
In particular, superconductivity 
is shown to disappear at any temperature when  $0.2\pi \lesssim \theta_3 \lesssim
\pi$ and $0.03 \lesssim h/E_F \lesssim
0.06$. Nonetheless, $\theta_3=0.5\pi$ and $1.5\pi$, overall,
provides smaller critical temperatures within $0.03 \lesssim h/E_F \lesssim
0.22$. 
These results demonstrate that the use of ferromagnets $\rm F_{2.3}$
with $h/E_F \ll 1$ causes the greatest  decline in the superconducting state.
Below, this will be discussed in terms of the population  of both
the equal-spin and opposite-spin triplet correlations in each relevant  region of the spin valve.
Lastly, in a similar fashion we 
show the importance of using half-metallic outer layers
to achieve enhanced spin valve effects.
In Fig.~\ref{ffsff_tc}(g), 
we exhibit 
 the
critical temperature 
vs the normalized exchange energy $h_4/E_F$.
 The ferromagnets  again 
have $D_{2,3}=10$ and
$h_{2,3}/E_F=0.05$.
The magnetization in $\rm F_2$ is aligned according to $\theta_2=0.5\pi$ (along the $y$-direction)
and the first outer layer has
$h_1=E_F$.
The exchange field strength in  $h_4$ varies from $0$ to $E_F$,
or equivalently, from a nonmagnetic normal metal to a 
fully spin polarized half-metal.
Consistent with
previous studies \cite{bernard1,half,singh}, 
the results shown here demonstrate  that 
when the layer adjacent to the weak ferromagnet 
has one spin state present  at the Fermi energy,
the greatest variations in $T_c$  as a
function of magnetization rotation can  occur. 
Switching between  superconducting and normal resistive states
has previously been found in \fsfz\cite{switch} and \sff\cite{half} structures.
By incorporating multiple half-metallic layers,
the variations in
$T_c$ found here, with a fairly thick S layer, are considerably
larger than what has been previously reported for the simpler \sff and \fsf counterparts. 

The conversion of opposite-spin triplet pairs into equal-spin triplet pairs
	is enhanced by coupling a weaker ferromagnet with the half-metal due in part
	to the preservation of phase coherence that would otherwise be destroyed by
	the single-spin half-metal.
To exemplify this, and  to compare
 the relative strengths of the
two types of spin valves, we present in Fig.~\ref{fig:comp}, 
 the normalized $T_c$ as a function of magnetization orientation $\theta_3$  for
  both the (a) $\rm F_1F_2SF_3$ and (b) $\rm F_1F_2SF_3F_4$ types of structures. 
  The thickness of the superconducting layer is fixed at $D_S=250$, which serves
  to effectively illustrate 
  which device configuration leads to the largest $\Delta T_c$ variations. 
  The ferromagnetic layers $\rm F_1$ and $\rm F_4$ are half metallic, 
  while the other layers are much weaker, standard ferromagnets. The remaining parameters are set identical to the
  cases previously shown.
 We examine three magnetization orientations in the $\rm F_2$ layer: $\theta_2=0, \pi/2, \pi$,
 while rotating $\theta_3$ continuously from $0$ to $2\pi$. 
 In the top panel, the angle 
 $\theta_2=0$ corresponds to a $\rm FSF$ configuration, 
 while for the lower panel it coincides with a $\rm F_1SF_2F_3$ structure.  
 Thus, by appropriately varying  
the magnetizations, the
multivalves can be reduced to their simpler 
 $\rm F_1F_2S$ and $\rm F_1SF_2$ spin-valve counterparts.   
 As seen, $\theta_2=0$ induces the weakest 
 variations in $T_c$ for both devices. 
 The case with $\theta_2=\pi$ introduces moderate
 variations in $T_c$, while the largest changes arise when 
$\theta_2=\pi/2$, corresponding to the 
spin multivalve configurations introduced in this paper. 
It is evident that the critical temperature for the \ffsff 
multivalve reaches zero when $\theta_2=\pi/2$ and $0.48\pi\lesssim\theta_3\lesssim 0.52\pi$, 
which is equivalent to a normal resistive state.   
Thus, it is apparent that the \ffsff device can provide stronger variations in the
superconducting critical temperature compared to
$\rm F_1F_2SF_3$ and simpler spin valves.


\begin{figure}[b!]
\includegraphics[width=8.7cm, height=7.30 cm]{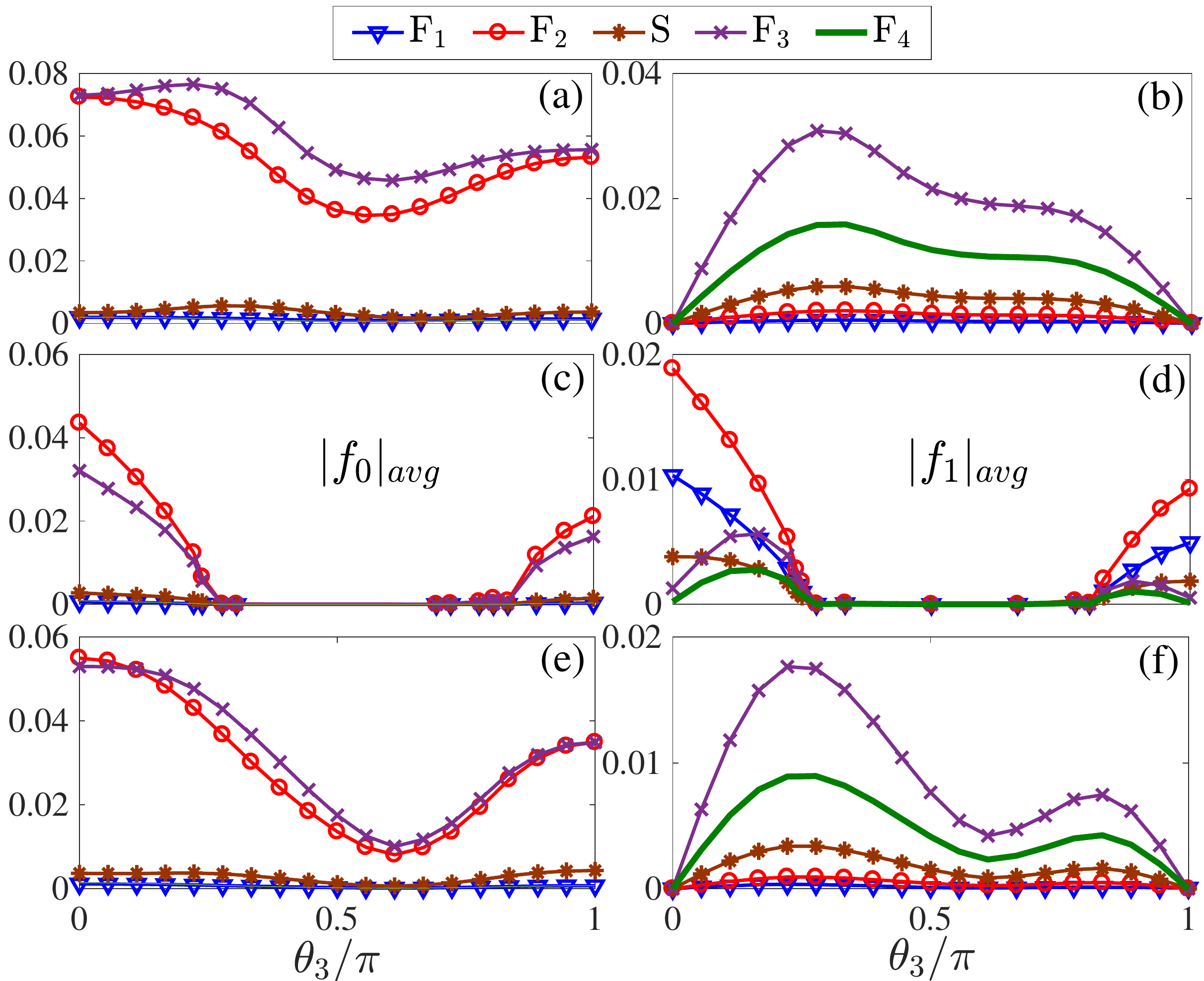}
\caption{The opposite-spin ($f_0$) and equal-spin ($f_1$) pairings vs $\theta_3$, averaged over each region in a \ffsff structure where
  F$_{1,4}$ are  half-metallic. In the top row, (a) and (b),
we set $\theta_2=0$, in the middle row, (c) and (d), $\theta_2=\pi/2$,
while in the bottom row, (e) and (f), $\theta_2$ is equal to $\pi$.}
\label{triplets}
\end{figure}

\begin{figure*}
\includegraphics[width=15.2cm]{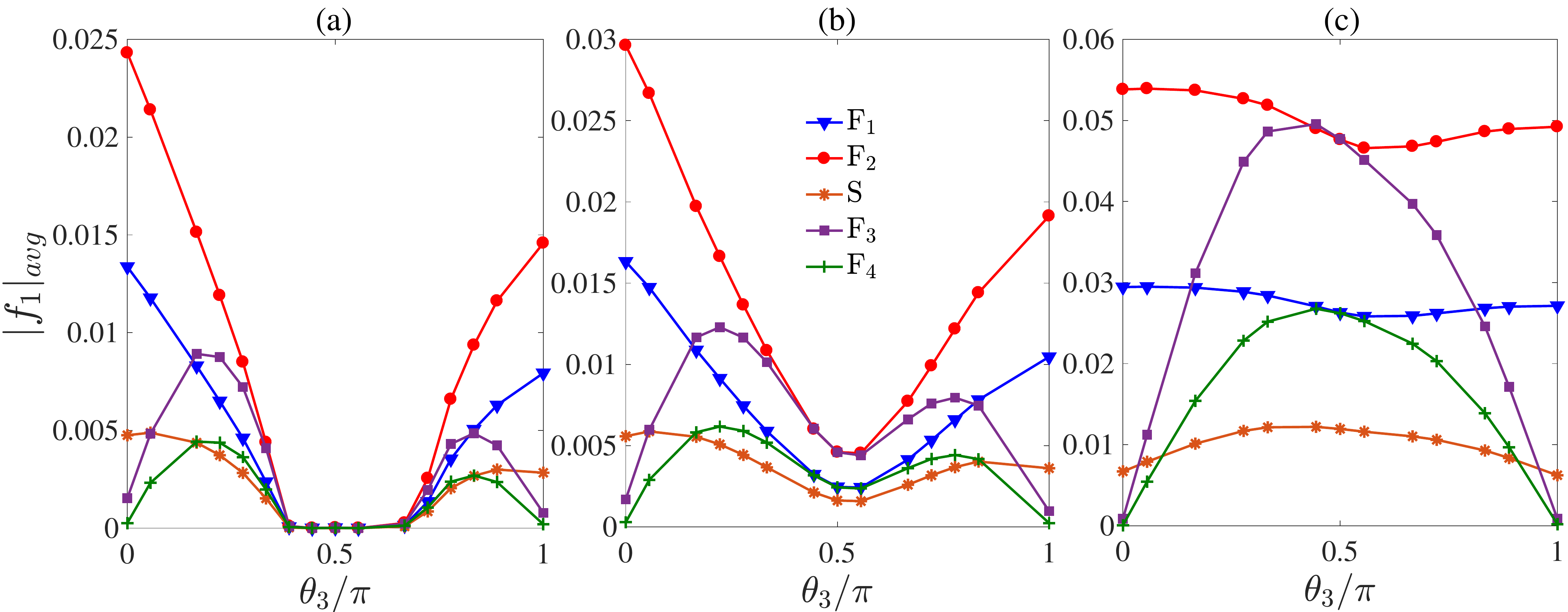}
\includegraphics[width=18.0cm]{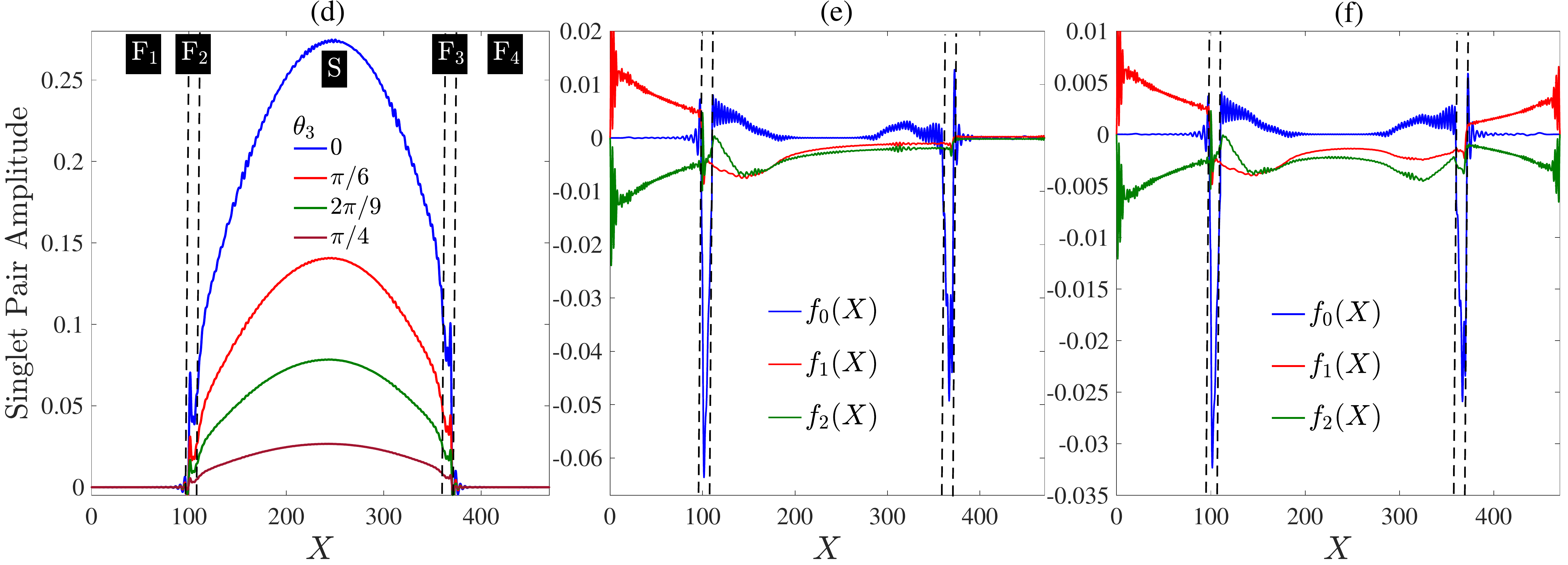}
\caption{Top row: Effects of the superconductor
width on the behavior of the equal spin triplets $f_1$ as
a function of the magnetization orientation angle $\theta_3$. Three cases are considered: (a) $D_S=260$,
(b) $D_S=270$, and (c) $D_S=400$.
The legend identifies each region of the \ffsffz\, multivalve in which $f_1$ is averaged over. Bottom row: The singlet (d) and triplet amplitudes ((e) and (f)) shown as a function of dimensionless 
position throughout the entire \ffsff spin valve. Here $D_S=250$, and $\theta_2=\pi/2$. 
The dashed vertical lines identify  the interfaces bounding the thin ferromagnets $\rm F_2$ and $\rm F_3$ (the regions are labeled in (d)). 
In (e) the magnetization orientation angle $\theta_3=0$, and in (f) $\theta_3=\pi/6$.
}
\label{corr_vs_X}
\end{figure*}

\subsection{Density of States}\label{dos}

The study of single-particle excitations in these systems
can reveal important signatures in the proximity
induced singlet and triplet pair correlations. A useful
experimental tool that probes these single-particle states
is tunneling spectroscopy, in which 
the local DOS, $N(x, \varepsilon)$, can be measured 
 as a function of position $x$ and
energy $\varepsilon$.
In Fig.~\ref{dos_b},  the local DOS is shown as a
function of the normalized quasiparticle energy $\epsilon/\Delta_0$
and normalized location $X$ within the F$_2$SF$_3$
region of a \ffsff
multivalve. 
All plots are normalized to the corresponding DOS in a
bulk sample of S material in its normal state.
To be
consistent, we use the same layer thicknesses and
exchange field magnitudes found in
Fig.~\ref{ffsff_tc}(a).
In Figs.~\ref{dos_b}(a)-\ref{dos_b}(e) the
magnetization in the F$_3$ layer is rotated
incrementally according to $\theta_3=0,
  \pi/3.6, \pi/2, \pi/1.65$, and $\pi$.
  The $\rm F_{1,2,4}$ layers have their magnetizations oriented along $z$,
 i.e.,  $\theta_{1,2,4}=0$. 
  We note that, as seen in Fig.~\ref{ffsff_tc}, the
  variation of both $\theta_2$ and $\theta_3$ results in a fairly wide
  range of magnetization directions where $T_c\sim 0$, and thus 
 in those cases the system is not superconducting at any temperature. 
  Hence,  $\theta_2=0$ is chosen so that $T_c$
  is nonzero over a wider range in parameter space. 
  In (a), the magnetizations in each layer are collinear and directed along $z$,
  allowing only the opposite-pair correlations to exist. Although there is no
  gap in the energy spectrum, near the center of the superconductor ($X\sim 225$),
traces of the BCS-like energy structure are seen. The self-consistent proximity effects
 within the vicinity of the interfaces (found near the endpoints
 of the $X$ range) 
 however result in an increase of 
 subgap states ($|\varepsilon/\Delta_0| <1$).
  We see in
  Fig.~\ref{ffsff_tc}(b) that the deviation 
  of the magnetization orientation in F$_3$ from
  0 to $\pi/3.6$ results in a peak at zero energy near the ${\rm S/F_3}$
  interface ($x\sim360$), and within  the S region. As
  Fig.~\ref{dos_b}(c) and \ref{dos_b}(d) depict,
  as the
  magnetization rotates closer to the $\pi/2$ orientation,
  this zero energy peak region becomes more localized and pronounced in
  energy, extending deeper into the S layer.
  Finally, in Fig.~\ref{ffsff_tc}(e),
  the relative 
  magnetizations are again collinear, with the magnetization in $\rm F_3$ 
  antiparallel to the $z$-axis ($\theta_3=\pi$),
  resulting in the disappearance of the zero energy mode and
  subsequent splitting into separate Andreev bound states. 
  Note that the microscopic method used here 
  accounts for the significant band curvature
  near the Fermi energy arising from the strong spin-splitting effects
  of the half-metallic
  layers, as evidenced by the particle-hole asymmetry in the DOS.\cite{char,active,half}

\subsection{Superconducting Correlations}\label{corr}

To correlate  the spin-triplet superconducting
correlations to the nonmonotonic behavior of $T_c$, we
plot the opposite-spin ($f_0$) and equal-spin ($f_1$) pair correlations in
Fig.~\ref{triplets}. We have 
averaged $f_0$ and $f_1$ over each region separately, identified  by 
F$_1$, F$_2$, S, F$_3$, and F$_4$. 
In this fashion, one can readily evaluate the spatial distribution of the 
different pairings, denoted by $|f_0|_{avg}$ and $|f_1|_{avg}$. 
We consider the 
\ffsff  configuration with the same parameter
values used in
Fig.~\ref{ffsff_tc}(a). 
In Figs.~\ref{triplets}(a) and
\ref{triplets}(b) we set $\theta_2=0$, in \ref{triplets}(c) and
\ref{triplets}(d) $\theta_2=\pi/2$, and in \ref{triplets}(e) and
\ref{triplets}(f) $\theta_2=\pi$. The outer half-metals also have
$\theta_1=\theta_4=0$. 
For the top and bottom rows when $\theta_2=0$, and $\theta_2=\pi$, respectively, 
all of the magnetic layers have collinear relative magnetizations (except
of course for $\rm F_3$, which has $\theta_3$ varying).
In these configurations, we see that 
$f_1$ is zero at $\theta_3=0$ and $\theta_3=\pi$,
since then, all layers are collinear with a single quantization axis
which prohibits the generation of
equal-spin triplet pairs.
We see that  $f_0$
 exhibits  the same  behavior as
 $T_c$
  at $\theta_2=0,\pi/2, \pi$,
i.e., decreases as $\theta_3$ approaches $\sim\pi/2$. 
The $f_1$ triplet 
pairing however increases simultaneously,  demonstrating a
direct link between the 
appearance of equal-spin correlations and a decrease
in $T_c$. 
In cases where $\theta_2=0$ and $\pi$, we see that the $f_1$
correlation has a large amplitude in the right half-metal
F$_4$. 
It is also evident that $f_0$ is negligible in the half-metallic regions $\rm F_1$ and $\rm F_4$, since
this opposite spin triplet channel is energetically unstable in the regions with only one spin band present.
The 
magnetization direction in F$_2$ results in drastic changes: As
can be seen in Figs.~\ref{triplets}(c) and \ref{triplets}(d), $f_1$
correlations penetrate all regions throughout the multivalve and both
$f_0$ and $f_1$ vanish within $0.35\pi\lesssim\theta_3\lesssim
0.75\pi$ which is consistent with the $T_c$ results where 
superconductivity disappears in the middle S layer.

The superconductor width plays a key role in the range of magnetization angles that
result in strong spin valve effects. In the top row of Fig.~\ref{corr_vs_X}, three larger S widths corresponding to 
$D_S=260, 270$, and $D_S=400$ are presented. Previously in Fig.~\ref{ffsff_tc}(a)
we found that for  $D_S=250$ and $\theta_2=\pi/2$, there was a sizable range of angles $\theta_3$
in which the singlet correlations vanished at $T=0$. In these regions of parameter space,
the triplet correlations must also vanish as demonstrated in
Fig.~\ref{triplets}(c) and (d). 
If the thickness of superconductor layer increases, the pair breaking effects of the surrounding magnets 
become less detrimental to superconductivity, and the angular window that 
results in the system transitioning to a normal resistive metal narrows and eventually disappears altogether.
For instance, in Fig.~\ref{corr_vs_X}(a), although the overall trends are the same,
the interval in which $f_1=0$ has been reduced to
$0.4\pi\lesssim\theta_3\lesssim 0.65\pi$  compared to the $D_S=250$ case in 
Fig.~\ref{triplets}(d). For $D_S=270$ in Fig.~\ref{corr_vs_X}(b), the triplet correlations $f_1$ now pervade every region
of the spin valve. Instead of vanishing over a certain range of magnetization orientation angles, 
$f_1$ dips to a minimum at $\theta_3 \approx \pi/2$ in every layer, including the superconductor. 
Finally, if the S layer is increased to $D_S=400$ as shown in (c), the behavior of $f_1$ changes drastically
throughout the system.
The  three layers consisting of S, $\rm F_3$ and the half-metal $\rm F_4$, which previously
had dips in 
$f_1$ at $\theta_3\approx\pi/2$, now have their situations reversed 
 so that
   the
 $f_1$
triplet population is enhanced  in those regions.
Thus for example, for thin S widths the $f_1$ correlations in the half-metal $\rm F_4$ are constrained to vanish over
a range of 
magnetization misalignment angles, including at $\theta_3=\pi/2$, and
as the S width increases, the constraint is lifted and what was once an 
absence or minimum of $f_1$, now peaks for the magnetically inhomogenous situation
$\theta_3\approx \pi/2$.

Further information can be gathered from the spatial dependence of the pair correlations.
In Fig.~\ref{corr_vs_X}(d), the singlet pair correlations are shown as a function of dimensionless position $X$
for several magnetization orientations $\theta_3$. The parameter values used here
are identical to those implemented  in
Figs.~\ref{triplets}(c) and (d), where $D_S=250$ and $\theta_2=\pi/2$.
Each curve represents a different magnetization orientation  described by the angle $\theta_3$ shown 
in the legend. As observed in Fig.~\ref{ffsff_tc}, $T_c$ exhibits considerable variations when
rotating from  $\theta_3=0$, where all magnetizations are aligned along the $z$ direction (except in $\rm F_2$,
which is directed along $y$),
to $\theta_3=\pi/2$, where the magnetizations between  the
ferromagnets and half-metals are orthogonal.
Indeed,
the transition temperature  rapidly diminishes  until eventually
the system transitions to the normal state for all temperatures. 
This is reflected in the 
self consistent singlet pair correlations at $T=0$ (Fig.~\ref{corr_vs_X}(d)), where the 
rapid decline in the S region
as a function of $\theta_3$
 is
clearly evident. The singlet correlations of course vanish in the half-metallic regions where only one spin state
is permitted.
Moreover, due to the asymmetric magnetization profile, the singlet profile is not symmetric,
exhibiting   a  greater  presence of singlet correlations in $\rm F_3$ compared to $\rm F_2$.
Due to the singlet-triplet conversion that takes place, we see a corresponding increase in the $f_0$ triplets
in $\rm F_2$ and decrease in $\rm F_3$ (panels (e) and (f)). Although the opposite-spin
triplet and singlet correlations cannot reside in the half-metallic regions,
in Fig.~\ref{corr_vs_X}(e), we see the presence of the equal-spin triplet components $f_1$ and $f_2$ in the half-metal ($\rm F_1$). Since these triplet pairs so not suffer from the energetically unfavorable Zeeman splitting, they can
subsist in the half-metal. Although $f_1$ and $f_2$ propagate throughout the entire $\rm F_1$ region with a slow 
spatial variation, these correlations are absent in the other half-metal $\rm F_4$ which has its magnetization
collinear with the adjacent ferromagnet.  
By rotating the magnetization in $\rm F_3$ however, Fig.~\ref{triplets}(b) shows that $f_{1,2}$ triplet pairs can be created 
in $\rm F_4$, peaking  when $\theta_3\approx \pi/6$. This is shown in detail in Fig.~\ref{corr_vs_X}(f) where
the equal-spin triplet pairs have also been amplified in the superconductor.

\section{conclusions}\label{con}
In summary, motivated by 
recent theoretical progress and  experimental advancements in 
superconducting spin valves, 
we have proposed  \ffsf and \ffsff  superconducting triplet spin 
multivalves that
host  multiple spin valve effects among adjacent F layers. 
We calculated the superconducting  transition
temperature, and the spatially-resolved density
of states vs the magnetization orientations and layer thicknesses. Our
results reveal that due to proximity effects and spin-valve effects involving  
singlet and triplet conversion and creation, these
structures offer stronger superconducting spin-switching and
spin-triplet generation compared to the 
basic \textit{single} \sff and \fsf spin-valve counterparts. 
 In order to provide  insight into these switching
effects and accurate details of the behavior of the pair correlations
in both the singlet and triplet channels, we performed our calculations 
using a microscopic self-consistent  theory that
is capable of handling the broad range of length and energy scales involved.
This method also allows for multiple Andreev reflections and
corresponding resonances in the ballistic regime where the mean free
path is much larger than the system thickness. 
Using this formalism, coupling between
layers and quantum effects arising from interfering quasiparticle 
trajectories within adjacent layers is accounted for. 
The results shown here demonstrated that 
the proposed hybrid structures can provide unambiguous signatures of the presence of equal-spin triplet
correlations that can arise under relatively weak in-plane external magnetic fields,
thus increasing device reliability.

\acknowledgments
K.H. is supported in part by ONR and a grant of HPC resources from the DOD HPCMP.

\appendix

\section{Spin Rotation}
\label{appA}
Here we outline the spin rotations 
involving 
the  triplet components ($f_0$, $f_1$, $f_2$),
affording a clearer physical interpretation of the results.
The central quantity that we use to perform the desired rotations
is the spin transformation matrix $\mathcal{T}$ in particle-hole
space. 
The quasiparticle amplitudes  transform as,
\begin{align}
\Psi^\prime_n (x)
 = \mathcal{T}
 \Psi_n (x), \label{transform}
\end{align}
where $\Psi_n (x)=(u_{n\uparrow}(x),u_{n\downarrow}(x),v_{n\uparrow}(x), v_{n\downarrow} (x))$,
and the prime denotes quantities in the rotated system.
The matrix 
$\mathcal{T}$  can be written 
solely in terms of the angles that describe the local
magnetization orientation.
In particular, when the orientation
of the exchange fields in a given layer
is expressed in terms of the angles 
given in Eq.~(\ref{fields}), we can write:
\begin{align}\label{tmatsmall}
\mathcal{T}= &\left[
  \begin{array}{cccc}
  \cos\left({\theta_i}/{2}\right)  &
    -i\sin({\theta_i}/{2}) & 0  &0\\
    -i\sin({\theta_i}/{2}) &
        \cos({\theta_i}/{2}) & 0 & 0\\
        0&0&\cos\left({\theta_i}/{2}\right)&
  -i\sin({\theta_i}/{2}) \\
  0&0& -i\sin({\theta_i}/{2})&
    \cos({\theta_i}/{2})
  \end{array}
\right].
\end{align}
Using the spin rotation matrix $\mathcal{T}$, it is also possible to transform the
original BdG equations ${\cal H}\Psi_n=\epsilon_n\Psi_n$ (Eq.~(\ref{bogo})) by performing the
unitary transformation: ${\cal H}' = \mathcal{T} {\cal H}
\mathcal{T}^{-1}$, with   $\mathcal{T}^\dagger \mathcal{T} =1$. 
As is the case under all unitary transformations,
the eigenvalues here
are preserved, but the eigenvectors are modified in general
according to Eq.~(\ref{transform}).
Thus we can write,
\begin{align}
u'_{n\uparrow}&=\cos\left({\theta_i}/{2}\right) u_{n\uparrow}-i\sin({\theta_i}/{2}) u_{n\downarrow}, \\
u'_{n\downarrow}&=\cos\left({\theta_i}/{2}\right) u_{n\downarrow}-i\sin({\theta_i}/{2}) u_{n\uparrow},\\
v'_{n\uparrow}&=\cos\left({\theta_i}/{2}\right) v_{n\uparrow}-i\sin({\theta_i}/{2}) v_{n\downarrow}, \\
v'_{n\downarrow}&=\cos\left({\theta_i}/{2}\right) v_{n\downarrow}-i\sin({\theta_i}/{2}) v_{n\uparrow}.
\end{align}
Thus for example,
the terms involved in calculating the singlet pair
correlations (Eq.~(\ref{sc})),
obey the following relation between the 
transformed (primed) 
and untransformed quantities: 
\begin{align}
{u'}_{n\uparrow} 
{v'}_{n\downarrow}^{*}+{u'}_{n\downarrow}{v'}_{n\uparrow}^{*} =
u_{n\uparrow} v_{n\downarrow}^* +u_{n\downarrow}
v_{n\uparrow}^*. 
\end{align}
Thus, the terms that dictate the
singlet pairing are invariant
for any choice of quantization axis, transforming as scalars under
spin rotations.

The terms governing the triplet amplitudes on the other hand are not invariant under spin-rotation. 
The relevant particle-hole products in Eq.~(\ref{f0})
that determine $f_0$, upon the spin transformations obey the following
relationships:
\begin{align}
{u'}_{n\uparrow} {v'}_{n\downarrow}^{*} -{u'}_{n\downarrow}
{v'}_{n\uparrow}^{*} &= \cos\theta_i (u_{n\uparrow} v_{n\downarrow}^{*}-u_{n\downarrow} v_{n\uparrow}^{*})\nonumber \\
&+i\sin\theta_i\left(u_{n\uparrow} v_{n\uparrow}^{*}-u_{n\downarrow} v_{n\downarrow}^{*}
\right), \nonumber \\
&=
f_0 \cos\theta_i 
+i\sin\theta_i f_2,
\end{align}
For the equal-spin component $f_1$, the rotation leaves $f_1'$ unchanged:
\begin{align}
&{u'}_{n\uparrow} {v'}_{n\uparrow}^{*} + {u'}_{n\downarrow} {v'}_{n\downarrow}^{*}
=
u_{n\uparrow} v_{n\uparrow}^*
+u_{n\downarrow} v_{n\downarrow}^{*}.
\end{align}
For $f_2'$ however, it is straightforward to show that
\begin{align}
{u'}_{n\uparrow} {v'}_{n\uparrow}^{*} -{u'}_{n\downarrow}
{v'}_{n\downarrow}^{*} 
&= \cos\theta_i (u_{n\uparrow} v_{n\uparrow}^{*}-u_{n\downarrow} v_{n\downarrow}^{*})\nonumber \\
&+i\sin\theta_i\left(u_{n\uparrow} v_{n\downarrow}^{*}-u_{n\downarrow} v_{n\uparrow}^{*}
\right),\nonumber \\
&=\cos\theta_i f_2
+i\sin\theta_i f_0,
\end{align}

\end{document}